\documentclass[prb,twocolumn,showpacs,aps]{revtex4}
\usepackage{color,soul}
\usepackage {graphicx}
\usepackage {epstopdf}

\begin{document}
\title{Effect of nonmagnetic Ti substitution on the structural, magnetic and transport properties in pyrochlore iridate Eu$_{2}$(Ir$_{1-x}$Ti$_{x}$)$_{2}$O$_{7}$}

\author{Sampad Mondal$^{a,b,c}$\footnote{Email:sampad100@gmail.com}, B. Maji$^d$, M. Modak$^b$, Swapan K. Mandal$^a$\footnote{Email:swapankumar.mandal@visva-bharati.ac.in} and S. Banerjee$^b$\footnote{Email:sangam.banerjee@saha.ac.in}}
\address{$^a$ Department of Physics, Visva-Bharati, Santiniketan 731235, India\\
	$^b$ Saha Institute of Nuclear Physics, 1/AF Bidhannagar, Kolkata 700064, India\\
	$^c$Ramsaday College, Amta, Howrah 711401, India\\
	$^d$ Acharya Jagadish Chandra Bose College, 1/1B, A. J. C. Bose Road, Kolkata 700020, India}
\begin{abstract}
	We have studied the effect of nonmagnetic Ti substitution Eu$_{2}$(Ir$_{1-x}$Ti$_{x}$)$_{2}$O$_{7}$ with the help of electrical transport and magnetic measurement. The minor structural modification enhances the orbital overlapping and favours its electrical transport properties with Ti doping though the tuning of SOC and U with site dilution opposes it. As a result, metal insulator transition (MIT) is disappeared and resistivity of the system throughout the temperature increases with Ti doping. The nature of the conduction mechanism at low temperature follows power law like variation. As the Ti$^{4+}$ is nonmagnetic, the introduction of Ti at Ir site dilutes the magnetic interaction at Ir octahedral network, which in turn decreases the magnetic moment and magnetic frustration in the system though the magnetic irreversibility temperature is hardly affected by Ti.
\end{abstract}
\maketitle
{\large\bf{Introduction}}
\vskip 0.2cm

Recently, in condensed matter physics 5d transition metal mainly iridium oxide have created much attention due to their exotic magnetic and electronic properties
originating from interplay between spin orbit coupling (SOC) and electronic correlation (U)\cite{Rau, Cao, Wan}. Pyrochlores offer a perfect illustration for exploring the interplay of
SOC and U in 5d transition metal oxide compounds due to their interpenetrating corner sharing tetrahedral structure tend to form a narrow flat band at Fermi level that enhance the effect of SOC and electron correlation energy, both the energy scale are comparable to the band width of the transition metal. Within the realm of 5d transition metal based pyrochlore compounds, particular attention is drawn to pyrochlore iridate R$_{2}$Ir$_{2}$O$_{7}$ (where R represents a rare earth element) owing to its captivating transport and magnetic characteristics, including phenomena such as spin liquid, spin glass, Weyl semimetal behavior, and metal insulator transitions.\cite{Zhang,Yanagishima,Matsuhira,Matsuhira1,Nakatsuji} In these compounds system goes from a metallic state for R= Pr to insulating state for R=Gd-Lu depending on the ionic radius of the rare earth atom and in the intermediate state shows a metal insulator transition for R=Nd-Sm.\cite{Matsuhira,Matsuhira1} As it is believed that there exist a topological phase near the MIT phase boundary, the family of pyrochlore iridate compound with R= Nd, Eu and Sm are the promising candidate to give exotic topological phase.\cite{Wan}

To analyse the physical property mostly for Ir lattice among the pyrochlore iridates our special interest on Eu$_{2}$Ir$_{2}$O$_{7}$ (EIO) as we can avoid f-d exchange interaction\cite{Chen} for nonmagnetic Eu. So this is the most preferable compound to study the role of SOC and electronic correlation associated with the system. The analysis of resistivity data reveals that EIO undergoes
from paramagnetic metal to Weyl semimetal (in an intermediate temperature range) and, finally, to an antiferromagnetic insulator at the lowest temperature.\cite{Tafti} Besides Terahertz optical conductivity measurement confirm the Weyl semimetal state in this compound.\cite{Sushkov} The ground state of the pyrochlore iridate system can be tuned by band filling of Ir 5d state via chemical substitution. Previous study have shown that different band filling in EIO compound gives various interesting magnetic and transport properties. The doping of Sr at Eu site in EIO gives non-Fermi liquid behavior above T$_{MI}$.\cite{Banerjee} Recently the Weyl semimetal like behavior is observed in the polycrystalline EIO compound by doping Bi at Eu site by Telang et al.\cite{Telang} The doping of Ca at Eu site in EIO gives doping induced metallic phase,\cite{Kaneko} although the doping of Cu at Ir site in EIO does not influence the MIT.\cite{Mondal}

Our current study is focused on exploring the interplay between SOC and U within the pyrochlore iridate system and we have examined the impact of substituting Ti$^{4+}$ ions at the Ir$^{4+}$ site. The minor structural change with doping enhances the transport properties but the tuning of SOC and U with site dilution, leads to contrary outcome, resulting in an overall increase in resistivity. The substitution of Ti$^{4+}$ (3d$^{0}$) at Ir$^{4+}$ (5d$^{5}$) site not only changes the SOC and U of the compound but also dilutes the magnetic structure of the system. 
\vskip 0.5cm
{\large\bf{Experimental details}} 
\vskip 0.2cm
All the polycrystalline samples Eu$_{2}$(Ir$_{1-x}$Ti$_{x}$)$_{2}$O$_{7}$, where $x$=0, 0.05, 0.1, 0.15 were prepared by solid state reaction method. High purity ingredient powder Eu$_{2}$O$_{3}$, IrO$_{2}$ and TiO$_{2}$ were mixed in stoichiometric ratio and grounded well. After pressing the mixture powder in pellet form, heated at 1273 K for 3 days with several intermediate grinding. All the samples were characterized by powder X-ray diffraction (XRD). The room temperature XRD measurement was taken by X-ray diffractometer with Cu K$_\alpha$ radiation. Structural parameters were determined using standard Rietveld refinement technique with Fullprof software package. XPS measurements at room temperature were carried out by using an Omicron Multiprobe Electron Microscopy System equipped with a monochromatic Al K$_\alpha$ X-ray source ($h\nu$ = 1486.7 eV). All the XPS spectra have been analyzed by using PeakFit software, where Shirley method was used for background subtraction. Magnetic measurement was taken by Superconducting Quantum Interference Device Magnetometer (SQUID-VSM) of Quantum Design in the temperature range 3 K - 300 K. Electrical and magnetic transport measurement were carried out by four probe method with temperature range 2 K-300 K using Physical Properties Measurement System (PPMS).
\vskip 0.5cm
{\large\bf{Result}} 
\vskip 0.2cm
\begin{figure*} 
	\centering
	\includegraphics[width= 14 cm]{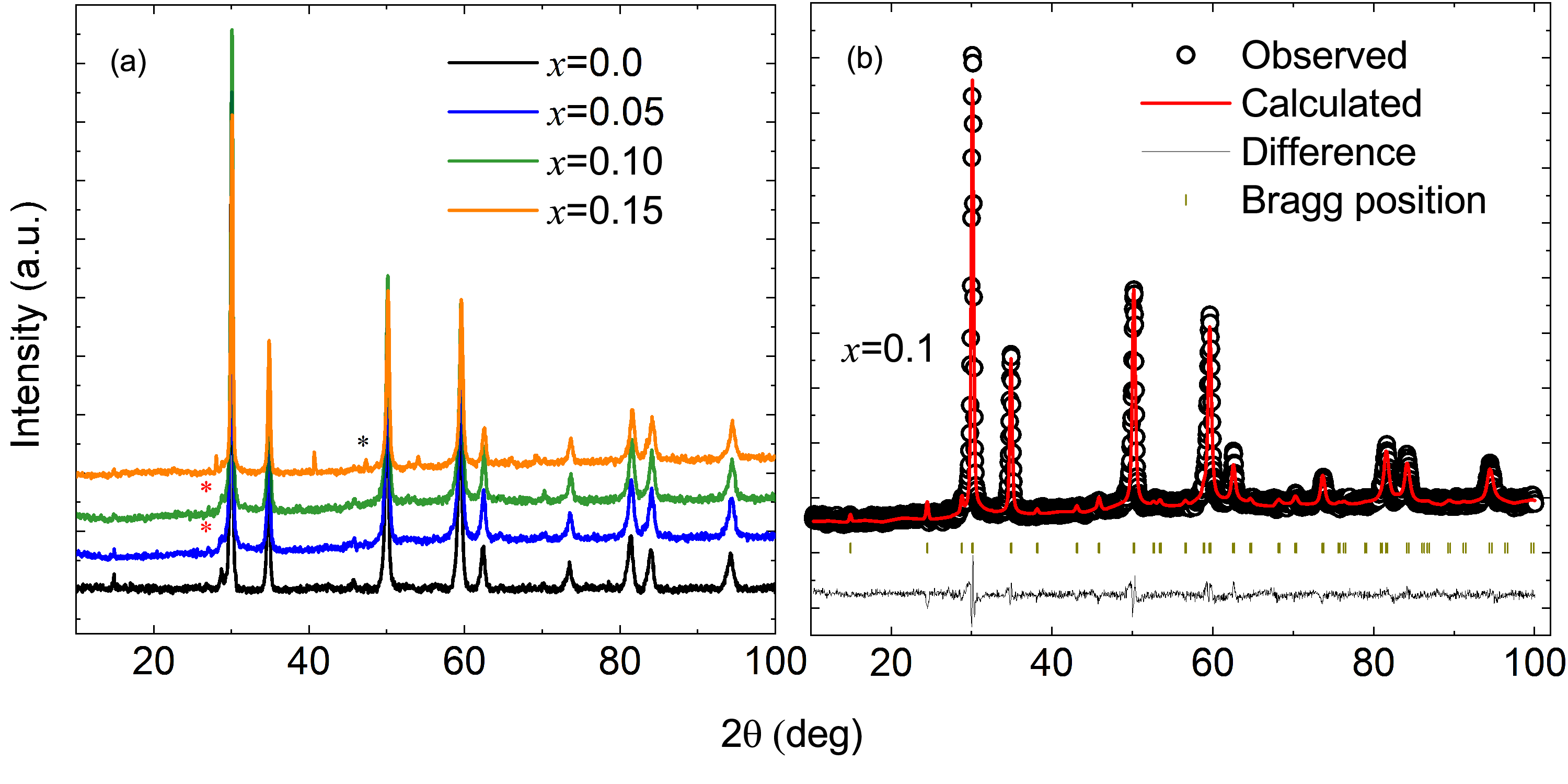}
	\caption{(a) Room temperature XRD patterns of the compounds Eu$_{2}$(Ir$_{1-x}$Ti$_{x}$)$_{2}$O$_{7}$, where $x=0, 0.05, 0.1, 0.15$ and red and
		black stars indicate the non reacting IrO$_2$ and Eu$_2$O$_3$ oxides respectively. (b) Rieteveld of the $x$=0.1 compound, where scattered data are observed data and red line is fit to the data.}                     
	\label{XRD_XRD_fit}
\end{figure*} 
\begin{figure*} 
\centering
\includegraphics[width= 14 cm]{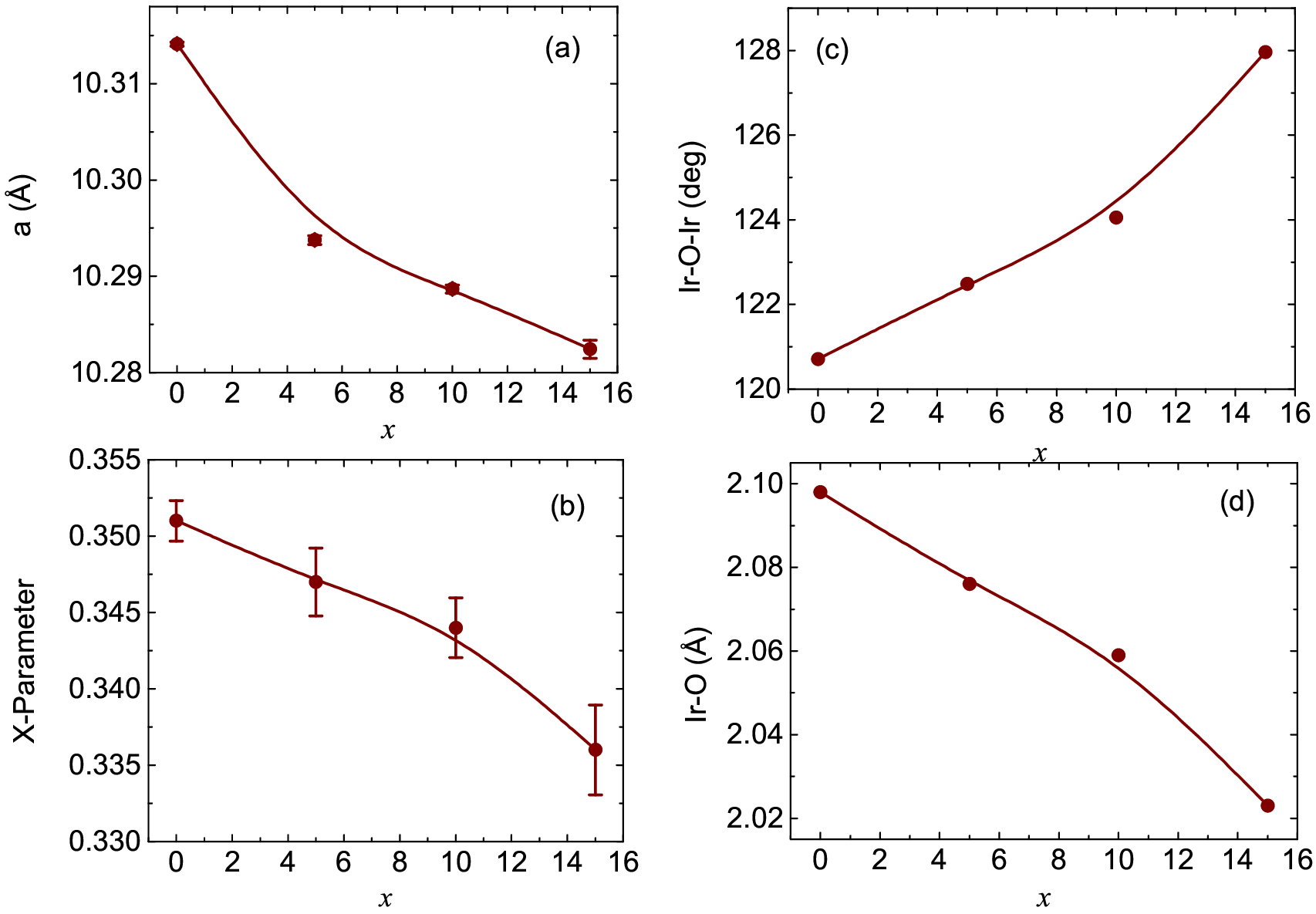}
\caption{(a) Variation of lattice parameter a (b) Positional parameter of O1 atom (X) (c) Ir-O-Ir bond angle and (d) Ir-O bond
	length with Ti concentration ($x$).}
\label{bond_inf}
\end{figure*}

Room temperature XRD patterns for all the compounds Eu$_{2}$(Ir$_{1-x}$Ti$_{x}$)$_{2}$O$_{7}$, where $x$=0, 0.05, 0.1, 0.15 are shown in fig. \ref{XRD_XRD_fit}(a). There is no significant change in the XRD pattern after doping, indicates that crystallographic structure and symmetry are retained with Ti doping. All the observed data are refined on the basis of cubic structure with space group Fd-3m. Fig. 1(b) shows the Rietveld refinement of the compound $x$=0.1. All the compounds are nearly in pure phase with the existence of some impurity phases. These phases occur due to nonreacting oxides IrO$_2$, Eu$_2$O$_3$, which are marked by red and black star respectively. The value of goodness of fit (defined as R$_{wp}/R_{exp}$) is found to be around 1.20, 1.06, 1.0, 1.54 for $x$=0, 0.05, 0.10 and 0.15 respectively, which indicates good fitting. Fig. \ref{bond_inf}(a) shows lattice parameter 'a' decreases with Ti doping systematically due to substitution of smaller ionic radius Ti$^{4+}$ (0.605 $\AA$) compare to Ir$^{4+}$ (0.625$ \AA$). In the pyrochlore structure two sets of oxygen atom present. Therefore the compound can be written as Eu$_{2}$Ir$_{2}$(O1$_{6}$)(O2). The crystallographic position of the four nonequivalent atoms are Eu at the 16d site ($\frac{1}{2}$, $\frac{1}{2}$, $\frac{1}{2}$), Ir at the 16c site (0, 0, 0), O1 at the 48f site (X, $\frac{1}{8}$, $\frac{1}{8}$) and O2 at the 8b site ($\frac{3}{8}$, $\frac{3}{8}$, $\frac{3}{8}$). Ir cation are coordinated with six O1 atoms and form corner shared IrO$_6$ octahedra, where each O1 atom is equal distance from Ir atom. Though position of the O1 atoms delivers distortion in the IrO$_6$ octahedra. For a perfect and undistorted octahedra, the value of the X is 0.3125 and Ir ions reside under the cubic field.\cite{Gardner} Deviation from the ideal value of X generates a distortion in the octahedra, which lower the cubic symmetry in the crystal structure. The value of X for parent compound is 0.352, which is larger than the ideal vale, suggesting a compressed and distorted IrO$_6$ octahedra. Fig. \ref{bond_inf}(b) shows the value of the X is decreasing with doping, suggesting the system headed toward perfect octahedra and gives rise to reduced crystal field. The bond angle Ir-O-Ir and bond length Ir-O related to IrO$_6$ octahedra have an impact role in the orbital overlapping and charge transfer mechanism, which contribute to the electronic properties of the system. Fig. \ref{bond_inf}(c-d) shows the doping dependents bond angle Ir-O-Ir and bond length Ir-O. Here bond angle is increasing with doping. When bond angle stretching toward 180$^\circ$, maximum electrons hopping between two Ir atom via O atom. Thus increasing of bond angle Ir-O-Ir delivers a increase of electron hopping. The decrease of bond length indicates the enhance of overlapping between Ir(5d) and O(2p) orbitals, which promotes larger hoping of Ir electrons. So increases of bond angle and decreases of bond length, both the structural modification enhance the orbital overlapping and favors its electrical transport properties.

\begin{figure} 
	\centering
	\includegraphics[width= 8 cm]{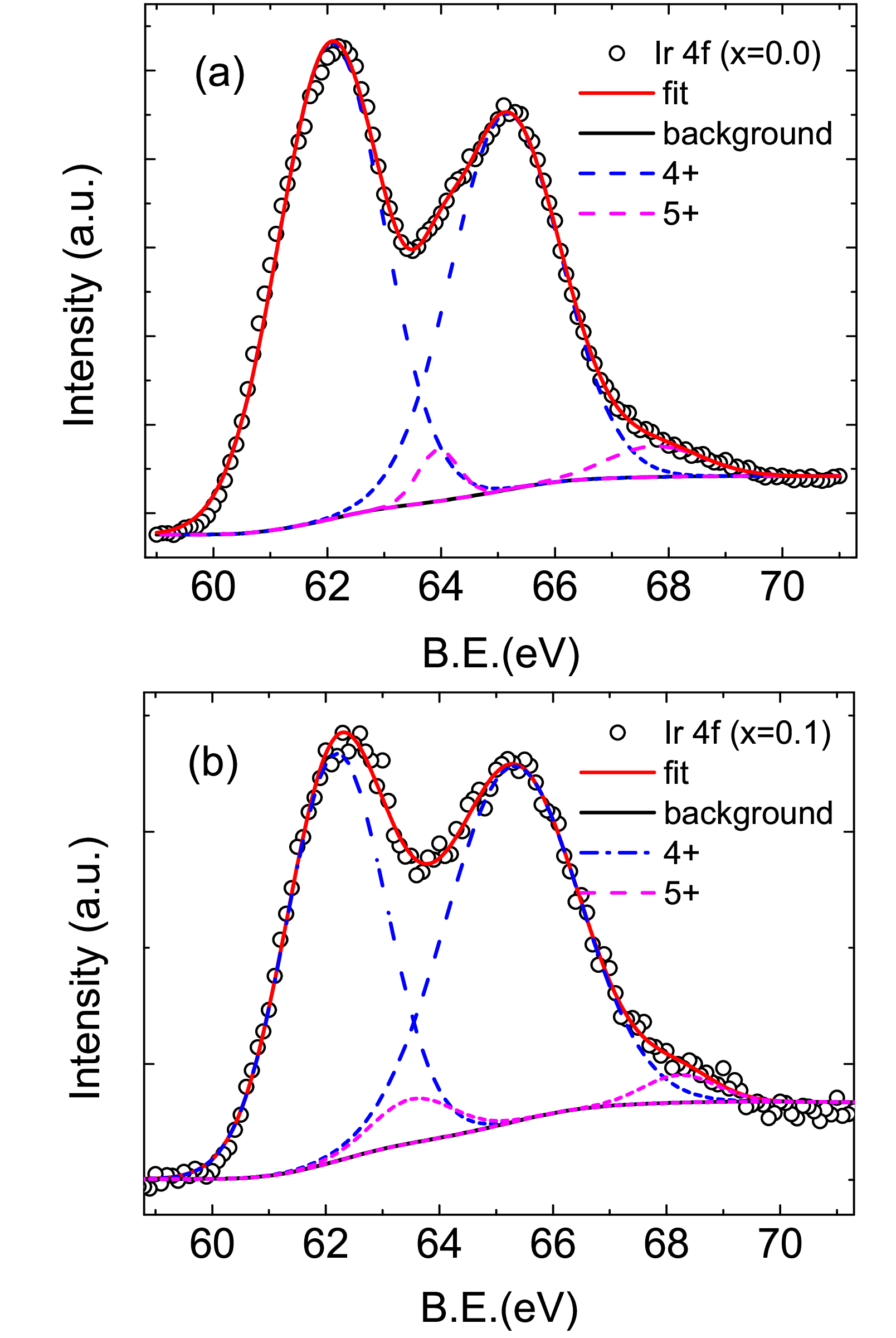}
	\caption{(a-b) show Ir 4f core level X-ray photoelectrons spectra for $x$=0 and 0.1.}                     
	\label{XPS}
\end{figure} 
In this system magnetic and transport properties are dominated by transition metal Ir. Therefore, it is important to know the charge state of the transition metal.
We have performed X-ray photoelectrons spectroscopy (XPS) measurement for the samples where fig. \ref{XPS}(a,b) depict the Ir 4f core level spectra for $x$=0, 0.1 respectively. Ir 4f core level spectra is split into 4f$_{7/2}$ and 4f$_{5/2}$ electronic states around the binding energy 62 and 65 eV respectively (as shown by dashed blue line) due to spin orbit coupling (SOC). All the spectra can be fitted by taking the peaks due to Ir$^{4+}$ with some weak additional peaks arising at the higher binding energy 64 and 67.7 eV (as shown by pink dotted line) which are correspond to 4f$_{7/2}$ and 4f$_{5/2}$ states of Ir$^{5+}$ ion. We observe the existence of the Ir$^{5+}$ peak for both the compounds. The calculation of area under the curve for XPS spectrum reveals that for $x$=0, 0.1 compounds, the amount of Ir$^{4+}$  and Ir$^{5+}$ are 95.9\% and 4.1\%, 96\% and 4\% respectively. Both the compounds contain near about equally amount of Ir$^{5+}$ state and majority of Ir is in Ir$^{4+}$ charge state. The small amount of Ir$^{5+}$ in both the compounds could be due to unavoidable nonstoichiometry of the sample.\cite{Ishikawa,Zhu}
\begin{figure} 
	\centering
	\includegraphics[width= 8 cm]{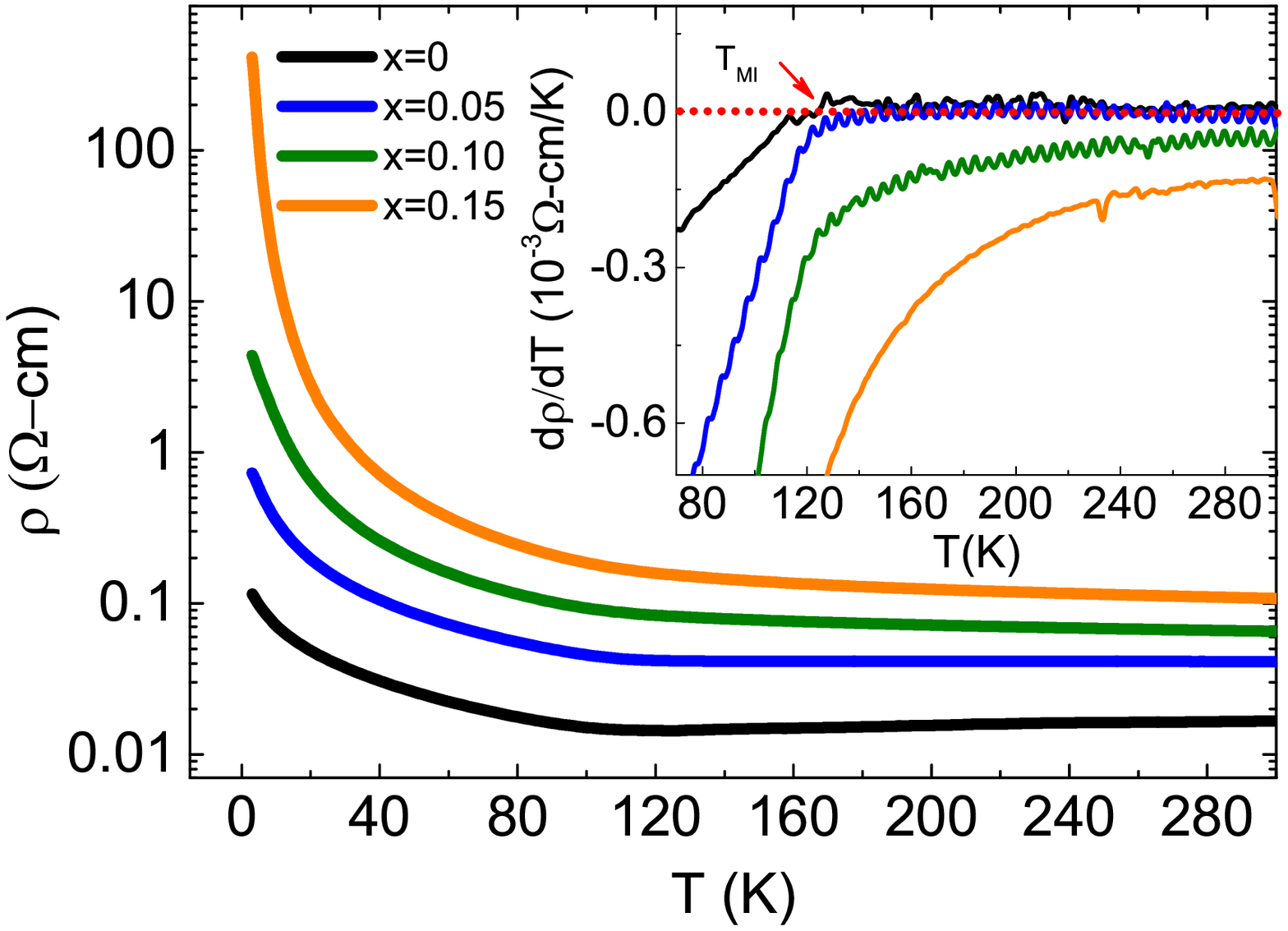}
	\caption{Temperature dependent resistivity of the compounds Eu$_{2}$(Ir$_{1-x}$Ti$_{x}$)$_{2}$O$_{7}$, where $x$=0, 0.05, 0.1, 0.15. Inset: Temperature derivative of the resistivity as a function of temperature.}                     
	\label{Rho-T}
\end{figure} 

Fig. \ref{Rho-T} demonstrates the temperature dependent electrical transport phenomena for all the samples. Parent compound shows metal insulator transition near about 120 K, below which the temperature dependent resistivity ($\rho$(T)) sharply increases with decrease in temperature. Such type of resistivity behaviour is previously reported.\cite{Ishikawa} The temperature derivative resistivity (d$\rho$/dT) changes sign from positive to negative upon cooling from room temperature for $x$=0 but does not change sign with Ti doping, shown in inset fig. \ref{Rho-T}. This indicates MIT is disappearing with Ti and tends the system into insulating state. $\rho$(T) for all the compounds increases systematically with Ti and at low temperature for the highest doped compound ($x$=0.15) increases four order of magnitude. This behaviour was also reported for a Ti doped layered Sr$_{2}$Ir$_{1-x}$Ti$_x$O$_{4}$, where $\rho$(T) increases systematically with Ti and end member Sr$_{2}$TiO$_{4}$ shows highly insulating nature.\cite{Gatimu,Ge}
\begin{figure} 
	\centering
	\includegraphics[width= 8 cm]{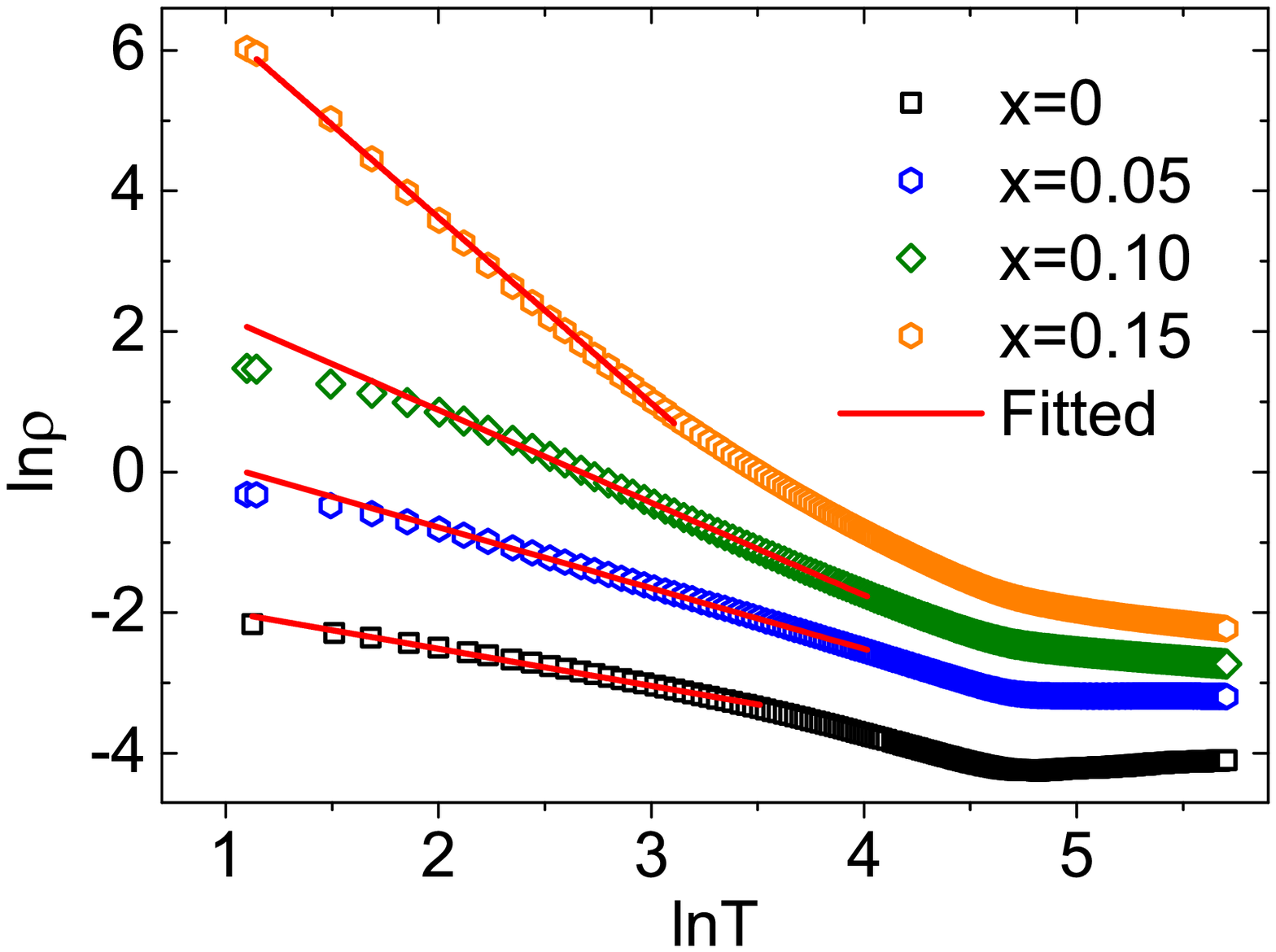}
	\caption{ln-ln plot of the $\rho$(T) data for all the compounds where red line is fit to the data.}                     
	\label{log-log}
\end{figure} 
\begin{table}
\centering \caption{Fitting parameters $\alpha$ obtained from Power law fitting for the samples Eu$_{2}$(Ir$_{1-x}$Ti$_{x}$)$_{2}$O$_{7}$ system.} 	
\begin{tabular}{ccc}
	\hline
	\hline Sample&\hspace{0.5cm}Temperature range (K)&\hspace{0.5cm}$\alpha$\\
	\hline $x$=0.0 & 3-23 & 0.53\\
	\hline $x$=0.05 & 3-55 & 0.86\\
	\hline $x$=0.10 & 3-55 & 1.32\\
	\hline $x$=0.15 & 3-20 & 2.64\\
	\hline
	\hline
\end{tabular}
\label{log-log_fittab_EITO}
\end{table}

To further investigate in more details, we have tried to fit the temperature dependent resistivity data at low temperature region with power law
\begin{equation}
	\rho = \rho_{0}T^{-\alpha}
\end{equation}
where $\alpha$ is the exponent, shown in fig. \ref{log-log}. The fitting temperature region and $\alpha$ value are reported in Table-\ref{log-log_fittab_EITO}. The parent compound fit within the temperature region 3-20 K and the value is 0.53, which was reported by Telang et al.\cite{Telang} For the doped samples range is 3-50 K except $x$=0.15 and $\alpha$ increases with Ti doping systematically. In a 3D disordered system Mott variable range hoping can be expected.
\begin{equation}
	\rho = \rho_{0}\hspace{0.2cm}exp((T_{0}/T)^{1/4})
\end{equation}
where,  T$_{0}$ = $\frac{21.2}{N(E_{F}) \xi^{3}}$, $N(E_F)$  and $\xi$ are density of states at Fermi level and localization length. Ishikawa et al. described the transport behaviour by using VRH model for single crystal sample.\cite{Ishikawa} We have also tried to fit the resistivity data with VRH model. But data fits only a limited temperature region (3-10 K). Polycrystalline sample have extrinsic scattering due to grain boundaries, which plays an important role in transport mechanism. Attempt to fit the data with Arhenious plot
\begin{equation}
	\rho = \rho_{0}\hspace{0.2cm}exp((E_{g}/KT))
\end{equation}
also fails, does not give the satisfactory result.
\begin{figure} 
	\centering
	\includegraphics[width= 8 cm]{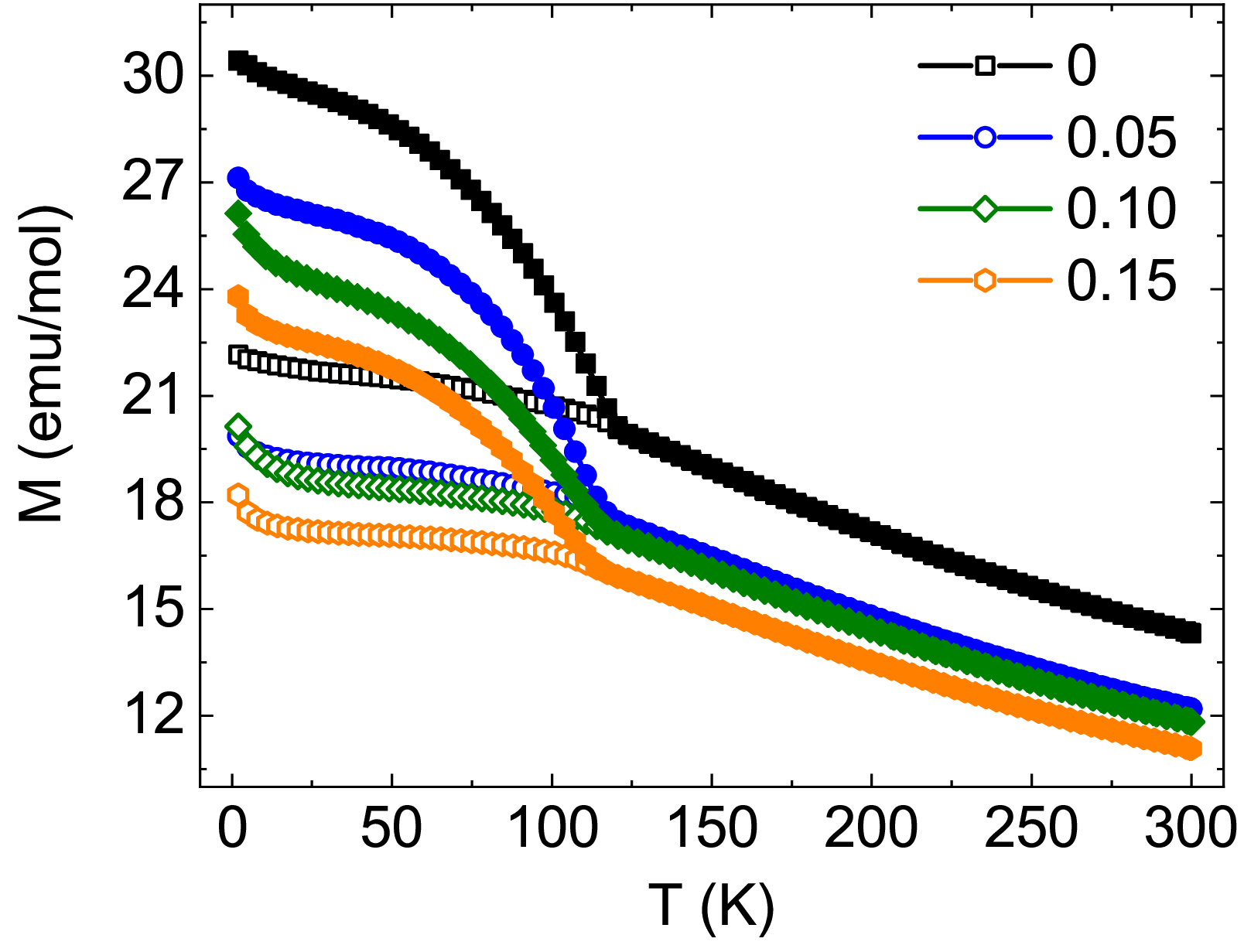}
	\caption{Temperature dependent magnetization measured at 1 KOe in ZFC FC protocol for  Eu$_{2}$(Ir$_{1-x}$Ti$_{x}$)$_{2}$O$_{7}$, where $x$=0, 0.05, 0.1, 0.15.}                     
	\label{M-T}
\end{figure} 

Fig. \ref{M-T} represent temperature dependent magnetization (M) for all the sample Eu$_{2}$(Ir$_{1-x}$Ti$_{x}$)$_{2}$O$_{7}$, where $x$=0, 0.05, 0.1, 0.15 under zero field cooled (ZFC) field cooled (FC) protocol with applied field 1 KOe. $x$=0 compound shows a magnetic irreversibility around T$_{irr}$ = 120 K, below which there is a bifurcation between ZFC and FC magnetization. The nature of the magnetic state for Eu$_{2}$Ir$_{2}$O$_{7}$ compound was well described in literature.\cite{Ishikawa} In this compound below  T$_{irr}$, Ir moments order in an antiferromagnetic all-in/all-out magnetic order, revealed by Resonant x-ray diffraction (RXD) and muon spin rotation and relaxation ($\mu$SR) measurements.\cite{Zhao,Sagayama} All the compounds show magnetic irreversibility and irreversibility temperature T$_{irr}$ for $x$=0, 0.05, 0.1, 0.15 compounds are 120, 118, 117, 115 K respectively. T$_{irr}$ shifting towards lower temperature systematically with Ti doping and below T$_{irr}$ the difference between ZFC-FC magnetization decreases with doping. We also observe that ZFC-FC curve shifting gradually towards the lower magnetization value with increasing of Ti concentration due to dilution of the magnetic lattice by nonmagnetic Ti$^{4+}$. The change of T$_{irr}$ with doping is not substantial where T$_{irr}$ decreases only by $\approx$ 5 K up to 15\% Ti doping. Such type of magnetic behavior was also observed for Cu and Ru doping at Ir site in Eu$_{2}$Ir$_{2}$O$_{7}$.\cite{Mondal,Wu} However, the hole doping at rare earth site in EIO significantly changes the irreversibility temperature with doping.\cite{Banerjee,Telang} Thus in these system the tuning of SOC and U by 3d/4d transition metal can play a role to minor change in irreversibility temperature. Besides, thermomagnetic irreversibility in a system depends on magnetic interaction, disorder and magnetic anisotropy. To check whether such effects are responsible for the minor change in irreversibility temperature further investigation is needed.


\begin{figure} 
	\centering
	\includegraphics[width= 8 cm]{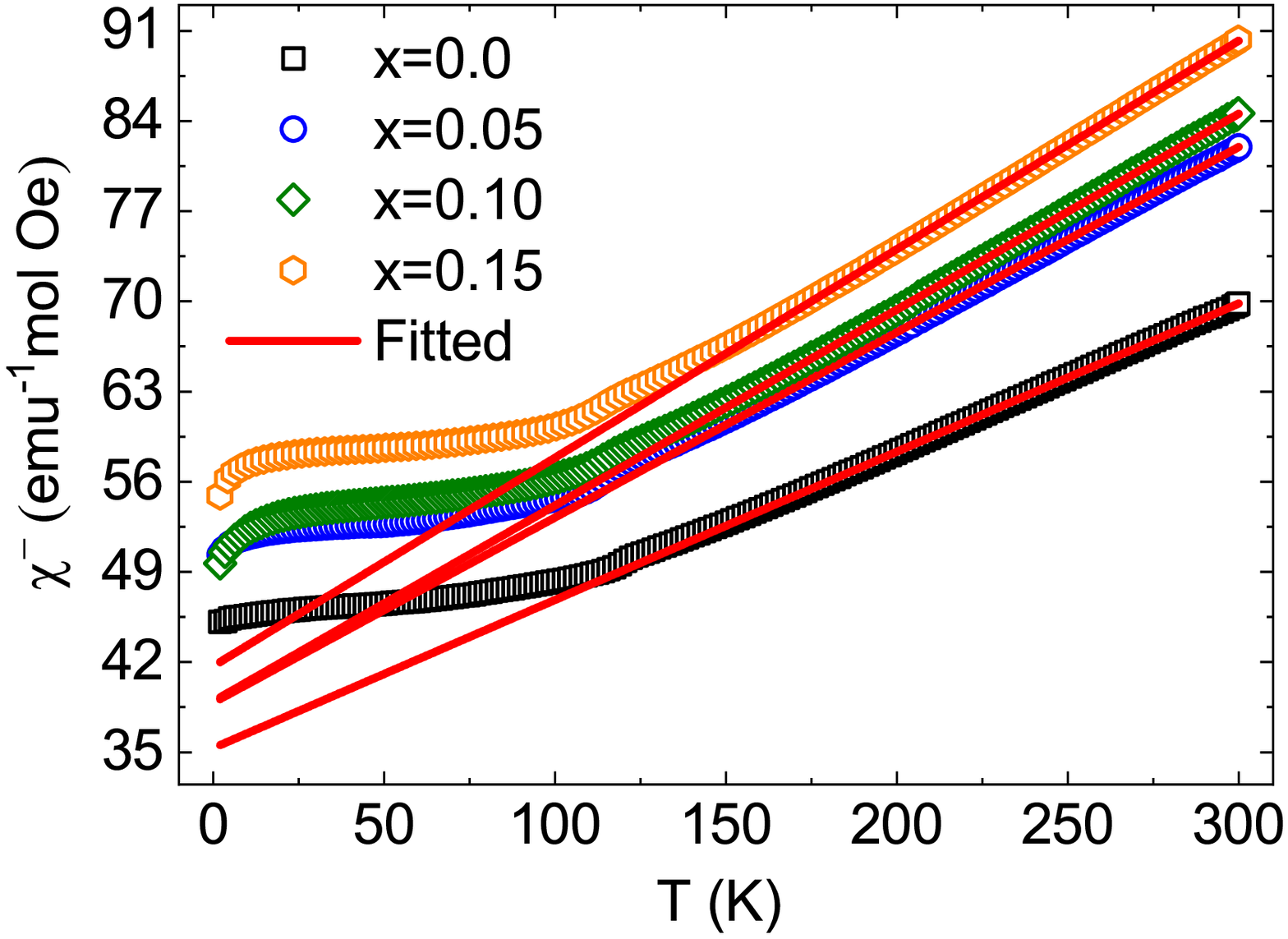}
	\caption{Temperature dependent inverse susceptibility with Curie fit for  Eu$_{2}$(Ir$_{1-x}$Ti$_{x}$)$_{2}$O$_{7}$, where $x$=0, 0.05, 0.1, 0.15  }                     
	\label{Invchi-T_EITO}
\end{figure}
Temperature dependent inverse susceptibility in the entire temperature range for all the compounds are shown in fig. \ref{Invchi-T_EITO}. In order to study the magnetic interaction, we have fitted the data in the temperature region 175-300 K with the Curie Weiss law $\chi = \frac{C}{T-\theta{p}}$, $\theta{p}$ is Curie Weiss temperature, and Curie constant $C=\frac{N_A\mu_{eff}^{2}}{3K_{B}}$, and  $\mu_{eff}$ is the effective paramagnetic moment. All the fitted parameters $\mu_{eff}$, $\theta_{p}$ and frustration parameter defined as $f = |\theta_p|/T_{irr}$ are shown in table-\ref{chifittab_EITO}. We observe negative $\theta_p$ for all the compounds, indicates the nature of the magnetic interaction in these compounds is antiferromagnetic, where strength of the interaction decreases with the introduction of Ti. Pyrochlore structure is inherently frustrated due to their geometrical structure, promotes the exotic ground states like spin glass,\cite{Gingras} spin liquid \cite{Nakatsuji,Gardner1} etc. The reported $f$ value for a highly frustrated pyrochlore oxide Y$_{2}$Mo$_{2}$O$_{7}$ is 8.8, gives a spin glass like behaviour.\cite{Gardner2} But in our system for the parent compound  $f$ value is lower than the highly frustrated system, which decreases with Ti. The effective paramagnetic moment $\mu_{eff}$ decreases with Ti due to incorporation of nonmagnetic Ti$^{4+}$ (3d$^{0}$) at Ir$^{4+}$ (5d$^{5}$) site.
\begin{table}
	
	\centering \caption{$\mu_{eff}$, $\theta_{p}$ and $f$ for Eu$_{2}$(Ir$_{1-x}$Ti$_{x}$)$_{2}$O$_{7}$ system.} 	
	\begin{tabular}{cccc}
		\hline
		\hline Sample&\hspace{0.5cm}$\mu_{eff}$ ($\mu_{B}$/f.u.)&\hspace{0.5cm}$\theta_{p}$ (K)& \hspace{0.5cm}$f$\\
		\hline x=0.0 & 8.33  &-306&2.55\\
		\hline x=0.05 &7.45 &-270&2.29\\
		\hline x=0.10 &7.25 &-260&2.22\\
		\hline x=0.15 &7.02 &-257&2.23\\
		\hline
		\hline
	\end{tabular}
	\label{chifittab_EITO}
\end{table} 
\begin{figure} 
	\centering
	\includegraphics[width= 8 cm]{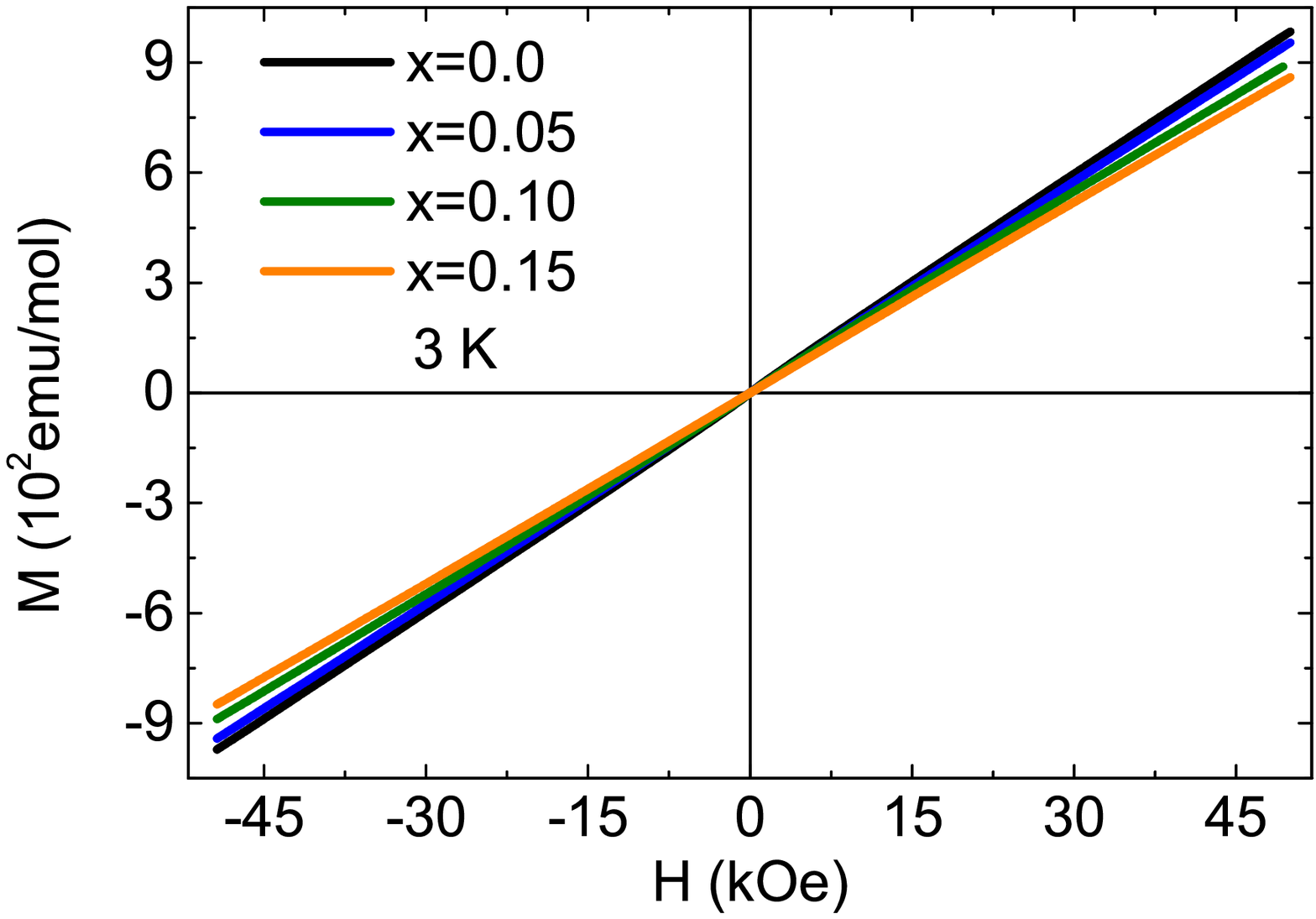}
	\caption{Isothermal magnetization at 3 K for Eu$_{2}$(Ir$_{1-x}$Ti$_{x}$)$_{2}$O$_{7}$, where $x$=0, 0.05, 0.1, 0.15 }                     
	\label{M-H}
\end{figure} 

Isothermal magnetization curve (M-H)for all the sample measured at 3 K up to field 50 kOe shown in fig. \ref{M-H}. Both the parent and Ti doped samples show linear behavior with field without any hysteresis. Magnetic moment up to highest applied magnetic field (50 kOe) decreases with Ti doping due to site dilution with nonmagnetic Ti$^{4+}$. 
\vskip 0.5 cm
{\large\bf{Discussion}}
\vskip 0.2cm
In Eu$_{2}$Ir$_{2}$O$_{7}$, most of the Ir is in the Ir$^{4+}$ (5d$^{5}$) charge state with small amount of Ir$^{5+}$ which is not changing with Ti doping. In this system structural arrangement of IrO$_6$ octahedra plays an important role in the transport and magnetic properties. The structural evaluation takes place with Ti doping. The doping of smaller ionic radius Ti at Ir site reduces the volume of the unit cell and also bond angle increases and bond length decreases. All the magnetic and transport properties of the system is governed by Ir 5d electrons, which have a strong spin orbit coupling and electronic correlation, both are comparable to the energy scale of Ir. In presence of IrO$_6$ octahedral environment, 5d orbital splits into t$_{2g}$ and e$_{g}$ energy level. It is believed that in presence of strong SOC further split the lower energy level t$_{2g}$ into spin J$_{eff}$=3/2 quadruplet with fully filled and spin J$_{eff}$=1/2 doublet with partially filled. Now if there exists on site electron correlation then a Mott like gap can open in the half filled J$_{eff}$=1/2 state, which makes the system J$_{eff}$=1/2 Mott insulator. As Ti being a 3d transition metal, it has higher electronic correlation (U) effect compare to 5d transition metal Ir. Thus the substitution of nonmagnetic Ti$^{4+}$ (3d$^{0}$) at Ir$^{4+}$ (5d$^{5}$) site not only dilute the magnetic network of the system but also tune its SOC and U. The analysis of the resistivity data, dilution of the Ir-O-Ir connectivity, tuning of the SOC and U and structural modification need to be considered. As the increase of bond angle Ir-O-Ir and decrease of bond length Ir-O enhances the orbital overlapping between Ir 5d and O 2p orbitals. This phenomena is likely to favour the electron transport mechanism. As a result resistivity was supposed to decrease with Ti doping. However, the site dilution and tuning of SOC and U by Ti doping may affect the electrical transport of the system. As a result the resistivity of the system increases with Ti doping as well as MIT is disappearing. $\rho$(T) also follows the power law like behaviour where power coefficient increases with Ti doping. The reason for the magnetic irreversibility is due to the field induced moment.\cite{Mondal} There is an unavoidable mixing of nonstoichiometry which produce nonmagnetic Ir$^{5+}$ state, deviates the all-in-all-out spin structure. With the application of field each octahedra picks up a moment, resulting as a field induced moment. As Ti$^{4+}$ (3d$^{0}$) is nonmagnetic, introduction of Ti at Ir site, Ti$^{4+}$ reduce the magnetic interaction in the octahedral network. As a result, magnetization of the system decreases with the incorporation of Ti at Ir site. T$_{irr}$ decreases very marginally with Ti doping. The tuning of SOC and U by Ti doping can play a role in minor change in irreversibility temperature. Besides various factors like disorder, magnetic interaction and anisotropy may affect the irreversibility temperature, further investigation needed. In this system the value of magnetic frustration parameter $f$ is lower than the highly frustrated pyrochlore oxide, which decreases with Ti due to magnetic site dilution by Ti$^{4+}$.
\vskip 0.5cm
{\large\bf{Conclusion}}
\vskip 0.2cm
From the analysis of electrical transport and magnetic property measurement in pyrochlore iridate Eu$_{2}$(Ir$_{1-x}$Ti$_{x}$)$_{2}$O$_{7}$, we conclude that with the introduction of nonmagnetic Ti influences the electrical transport and magnetic property of the system. Minor structural changes due to Ti doping enhances the orbital overlapping between Ir(5d) and O(2p) orbitals and favours the transport mechanism. But, as the $3d$ transition metal Ti has higher electron correlation than  $5d$ transition metal Ir, the introduction of Ti at Ir site tune the SOC and U of the system and site dilution occurs. As a result resistivity of the system overall increases systematically with the incorporation of Ti. The nonmagnetic Ti$^{4+}$ sits at the vertices of the Ir octahedra, dilute the magnetic interaction between Ir$^{4+}$. We find that magnetic moment and magnetic frustration decreases with Ti though magnetic irrevesibility temperature is hardly influenced by Ti.
\vskip 0.5cm
{\large\bf{Acknowledgements}}
\vskip 0.2cm
S. M. acknowledges Prof. S. Banerjee and Dr. S. K. Mandal (Supervisor) for their valuable contribution. S. M. would like to thank Prof. S. Hazra, Mr. G. Sarkar, SPMS Division, SINP for XPS measurement and Dr. M. K. Mukhopadhyay, Dr. R. Dev Das, SPMS Division, SINP for XRD measurement. This work is partially supported by SERB, DST, GOI under TARE project (Grant No.
TAR$/2018/000546$).

\end{document}